# Evaluation of Rail Decarbonization Alternatives: Framework and Application

**Adrian Hernandez**[†]
Graduate Research Assistant
Department of Civil and Environmental Engineering
Northwestern University
2145 Sheridan Road
Evanston, IL 60208, USA
Email: adrianhernandez2025@u.northwestern.edu

**Max T.M. Ng**[†]
Graduate Research Assistant
Transportation Center
Northwestern University
600 Foster Street
Evanston, IL 60208, USA
Email: maxng@u.northwestern.edu

**Choudhury Siddique**
Transportation Energy Analyst
Energy Systems and Infrastructure Analysis
Argonne National Laboratory
9700 S Cass Ave
Lemont, IL, 60439, USA
Email: csiddique@anl.gov

**Pablo L. Durango-Cohen**
Associate Professor
Department of Civil and Environmental Engineering
Northwestern University
2145 Sheridan Road
Evanston, IL 60208, USA
Email: pdc@northwestern.edu
Tel: 847.491.4008

**Amgad Elgowainy**
Senior Scientist and Distinguished Fellow
Energy Systems and Infrastructure Analysis
Argonne National Laboratory
9700 S Cass Ave
Lemont, IL, 60439, USA
Email: aelgowainy@anl.gov

**Hani S. Mahmassani***
William A. Patterson Distinguished Chair in Transportation
Director, Transportation Center

Northwestern University
600 Foster Street
Evanston, IL 60208, USA
Email: masmah@northwestern.edu

**Michael Wang**
Interim Division Director for Energy Systems and Infrastructure Analysis
Argonne National Laboratory
9700 S Cass Ave
Lemont, IL, 60439, USA
Email: mwang@anl.gov

**Yan (Joann) Zhou**
Interim Center Director for Systems Assessments
Energy Systems and Infrastructure Analysis
Argonne National Laboratory
9700 S Cass Ave
Lemont, IL, 60439, USA
Email: yzhou@anl.gov

† Co-First Author
* Corresponding Author

**ABSTRACT**

The Northwestern University Freight Rail Infrastructure & Energy Network Decarbonization (NUFRIEND) framework is a comprehensive industry-oriented tool for simulating the deployment of new energy technologies including biofuels, e-fuels, battery-electric, and hydrogen locomotives. By classifying fuel types into two categories based on deployment requirements, the associated optimal charging/fueling facility location and sizing problem are solved with a five-step framework. Life cycle analyses (LCA) and techno-economic analyses (TEA) are used to estimate carbon reduction, capital investments, cost of carbon reduction, and operational impacts, enabling sensitivity analysis with operational and technological parameters. The framework is illustrated on lower-carbon drop-in fuels as well as battery-electric technology deployments for US Eastern and Western Class I railroad networks. Drop-in fuel deployments are modeled as admixtures with diesel in existing locomotives, while battery-electric deployments are shown for varying technology penetration levels and locomotive ranges. When mixed in a 50% ratio with diesel, results show biodiesel's capacity to reduce emissions at 36% with a cost of $0.13 per kilogram of $CO_2$ reduced, while e-fuels offer a 50% emissions reduction potential at a cost of $0.22 per kilogram of $CO_2$ reduced. Battery-electric results for 50% deployment over all ton-miles highlight the value of future innovations in battery energy densities as scenarios assuming 800-mile range locomotives show an estimated emissions reduction of 46% with a cost of $0.06 per kilogram of $CO_2$ reduced, compared to 16% emissions reduction at a cost of $0.11 per kilogram of $CO_2$ reduced for 400-mile range locomotives. The NUFRIEND framework provides a systematic method for comparing different alternative energy technologies and identifying potential challenges and benefits of their future deployments.

**Keywords:**

freight, battery, biofuel, life cycle analysis, techno-economic analysis

# 1   INTRODUCTION

The transportation sector is the largest contributor to greenhouse gas (GHG) emissions in the US, contributing 27% of the emissions in 2020 (*1*). Many transportation modes, particularly in the freight sector, have been difficult to decarbonize due to their massive energy requirements and the associated investments that would be necessary for that purpose. However, recent advances in lower-carbon fuels, battery technology, and hydrogen fuels have provided potentially viable alternatives to diesel for these traditionally hard-to-decarbonize modes.

In 2019, the US freight rail sector accounted for approximately 40% of the national freight ton-miles and emitted nearly 40 megatons of $CO_2$ into the atmosphere in the process, an amount equivalent to the emissions of all the passenger vehicles in Texas alone (*2*, *3*). Though freight rail offers about four times greater energy efficiency than trucking (*4*), recent strides in the electrification of trucks (*5*) may significantly reduce rail's environmental advantage and cause freight demand to shift away to less energy efficient modes. As rail freight's importance in the overall supply chain continues to grow in the era of e-commerce (*6*), freight demand is forecast to grow rapidly in the coming decades (*7*), which may counteract railroads' investments in engine efficiency improvements. External pressures have also been mounting to decarbonize freight rail as local governments have considered regulations on locomotive idling in urban areas (*8*) and large shippers such as Amazon and IKEA have committed to net-zero carbon emissions by 2040 which include those produced by the shipment of their goods (*9*).

Diesel-electric locomotives have dominated US freight rail operations since the 1960's (*10*) and have seen significant improvements in powertrain efficiencies since that time (*11*). With the exception of a few corridors in the Northeast, track electrification has been limited to passenger rail as it would place a significant economic burden on private freight railroads to deploy electrical infrastructure in mostly rural stretches of the country and upgrade the many track segments that cannot accommodate overhead rail due to height constraints (*12*). Advancements in alternative energy storage technologies in recent decades—particularly in lower-carbon drop-in fuels, battery chemistries, and cleaner hydrogen pathways—offer a practical alternative to track electrification for decarbonization. Railroads and fuel chemists now have a larger portfolio of lower-carbon diesel replacements (e.g., biodiesel, electric-fuels, renewable-diesels) than they did a decade ago (*11*). Innovations in battery chemistry have led to increased volumetric and gravimetric energy densities, while reducing their overall cost per energy storage capacity (*13*), making this technology sufficiently mature to power electric locomotives (*14*). Hydrogen combustion and fuel cell experimentation has made the technology viable for locomotive applications (*11*), while experimentation in fuel production has yielded many different kinds of hydrogen fuel pathways (e.g., steam-methane reforming, electric, nuclear, renewable), each with differences in their environmental impacts and costs of production (*15*). Each of these alternative technologies provide distinct benefits and challenges to their implementation and must be compared on the economic, environmental, and operational impacts of their deployment to appropriately assess their value.

Several high-profile pilot studies have been conducted in partnership between multiple railroads, locomotive manufacturers, and local and state governments to test the viability of alternative technologies on revenue service (*11*, *16*, *17*). The 2019 BNSF-Wabtec battery-electric pilot ran a battery-electric locomotive in a diesel-hybrid consist on revenue service between the 300-mile

Stockton-Barstow route in California, showing emissions reductions of approximately 15% (*16*). In partnership between the Pacific Harbor Line and Progress Rail, a battery-electric switcher locomotive was run in the Port of Los Angeles and Long Beach to investigate its performance while reducing carbon emissions and eliminating all localized pollutant emissions (*17*). The Union Pacific Railroad has purchased 20 battery-electric locomotives for use as yard switchers, making it the largest commercial investment in the technology to date (*18*). After running a hydrogen fuel cell locomotive pilot, Canadian Pacific has committed to expanding its fleet of hydrogen locomotives and constructing two hydrogen production facilities to supply their operations (*19*).

Picking the right mix and schedules to invest and deploy the next-generation of energy technologies is a challenging process. Technological uncertainties, network effects, regional economics, and economies of scale all render mathematical optimization formulations of the problem essentially intractable. Decarbonization decisions will no-doubt have far-reaching environmental, operational, and financial impacts on railroads, shippers, regulators, and other stakeholders in the greater supply chain. While previous research focused on conventional fuel types and highly simplified railroad networks, there is a significant research gap in developing optimization models to support the deployment of infrastructure to support rail decarbonization.

## 1.1   Research Aims and Contributions

The Northwestern University Freight Rail Infrastructure & Energy Network Decarbonization (NUFRIEND) Framework was developed to assist the rail industry in planning and evaluating the adoption of alternative fuels for decarbonization efforts. Scenario-specific simulation and optimization modules provide estimates for emissions reduction, capital investments, cost of carbon reduction, and operational impacts for any deployment profile.

The framework relies on two different approaches to capture the characteristics and requirements of the two main groups of energy technologies:

1. **Drop-in fuels (Figure 1)**: Lower-carbon drop-in fuels can directly replace diesel fuel in locomotives and refueling stations. We assume no significant changes to existing assets or infrastructure are required for their deployment. Figure 1 shows the flowchart of the framework developed to analyze drop-in fuel deployment scenarios.

2. **Energy storage fuel technology (Figure 2)**: Hydrogen and battery-electric technology deployment poses a more complex problem as they require significant investments to be made in the siting of refueling/charging facilities and the replacement of locomotive fleets. Thus, strategies for locating and sizing refueling/charging facilities on a railroad's network to meet their energy demands must be developed to aid deployment decisions. However, jointly locating and sizing facilities quickly becomes a combinatorial problem due to the interconnectivity of the many potential facility locations seen on networks as well as the fact that regional economics and economies of scale both affect the cost of a facility deployment strategy. To reduce the problem complexity, we decompose the facility location and facility sizing problems from each other and insert a flow routing module in between to assign freight flow that must be served by the selected facilities. Figure 2 depicts a flowchart highlighting the five-step framework developed to address the deployment of the refueling/charging infrastructure to support hydrogen or battery-electric locomotives.

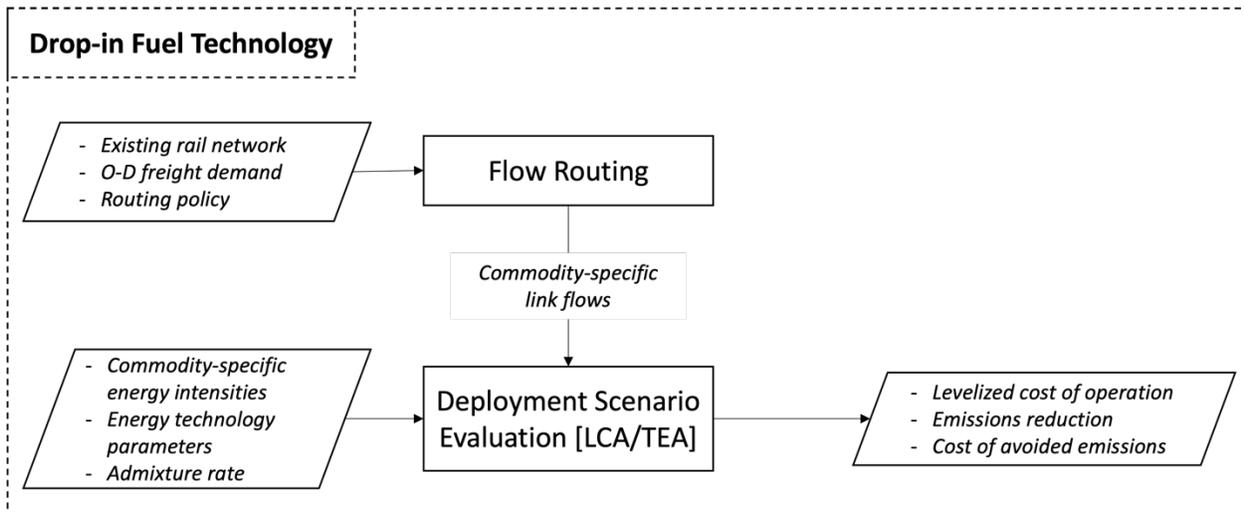

**Figure 1 - Flowchart of framework to support the deployment of lower-carbon drop-in fuels on the rail network.**

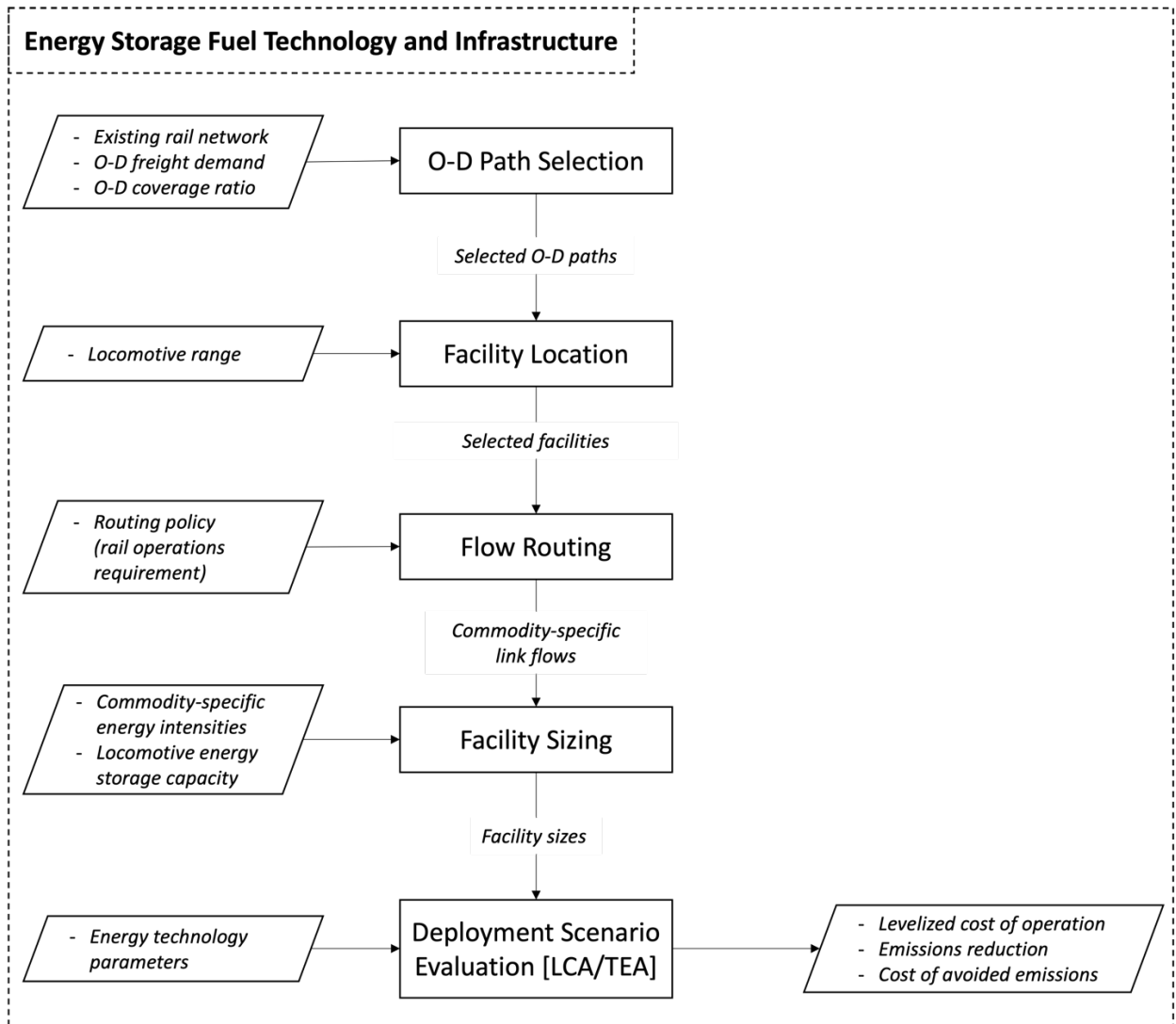

**Figure 2 - Flowchart of five-step framework to support the deployment of refueling/charging infrastructure for hydrogen and battery-electric technologies.**

The NUFRIEND framework aids stakeholders in analyzing alternative fuel technology deployments for freight rail operations. We assess and classify potential energy technologies based on deployment requirements and provide practical alternatives to diesel locomotives. The otherwise intractable facility location and sizing problem are solved with a five-step framework consisting of nominal problems from graph theory. A key advantage is the flexibility to apply the framework for any railroad considering the specific network structure and freight demand, outputting evaluation metrics for the associated emissions and costs relative to diesel operations. Equipped with the capability to efficiently simulate technology adoption scenarios with life-cycle analysis (LCA) and techno-economic analysis (TEA), this framework supports sensitivity analyses for different operational and technological parameters through a transparent and flexible input module, thereby addressing the uncertainties surrounding technological developments.

This paper reviews the relevant literature in freight rail decarbonization and new energy technologies in Section 2. The methodology of the five-step NUFRIEND framework, in particular

LCA and TEA, is discussed in Section 3, before introducing the data sources and parameters adopted in Section 4. The results are then illustrated with examples and scenario comparisons in the US freight rail network in Section 5, concluding with a discussion of limitations and future research directions in Section 6.

## 2    LITERATURE REVIEW

Network design and network flow optimization problems have seen rich applications in rail, with a review of recent work on their applications in multimodal freight in (*20*). Dynamic traffic assignment problems for freight movements have been studied for national networks (*21*), with extensions considering link congestion impacts on travel times (*22, 23*). Work to streamline rail operations by reducing fuel consumption and increasing system efficiency has focused on both emissions and cost reduction objectives (*24, 25*).

Early approaches to conducting freight rail decarbonization analyses date back to the 1970's with (*26, 27*) which analyzed the viability of track electrification, albeit from a primarily economic perspective, excluding environmental impact analyses. More recently, track electrification for the US freight rail network is considered as a budget-constrained network deployment problem with a built-in dynamic traffic assignment module to model the effects of track electrification on network congestion when the overall objective is to reduce rail emissions (*28*).

The analysis of decarbonization pathways such as hydrogen or battery-electric technologies require the deployment of refueling/charging facilities over the rail network. Taxonomies of facility location problems appear in (*29–31*). Such models address applications in warehousing and logistics (*32–34*) and electric vehicle charging station deployment (*35, 36*), but do not consider how the deployment of facilities affects routing or capacity decisions and the resulting cost and emissions impacts.

Previous research efforts evaluate fuel efficiency and resulting emissions mainly through system-average efficiency factors (*37, 38*), micro-simulation (*39*), mesoscopic simulation (*40*) and intermediate regression (*39, 41, 42*). Elgowainy et al. (*43*) updates Argonne National Laboratory (ANL)'s system-wide average efficiency factors using operational data from railroads, which are further updated with commodity-specific efficiency estimates in GREET (*15*). Heinold (*44*) compares five emissions estimation models and identifies each method's sensitivity to parameters such as number of rail cars, payload, speed, distance, and number of stops. Fullerton et al. (*41*) further suggests effects of trips, train/fuel, rail physics, locomotive mechanics, and topography on fuel efficiency. ICF International (*45*) evaluates the rail emissions associated with the haulage of different rail car types. Our framework goes beyond the descriptive nature of prior work in the transportation emissions space and applies these tools to evaluate and compare different alternative technology deployment scenarios, which can be used to guide investment and policy decisions.

As diesel-electric locomotives combust diesel to generate electricity onboard to power an electric motor, battery tender cars can be attached to power the electric motor directly with limited retrofitting (*46*). An economic and environmental analysis of battery tender car attachments shows that under the current electricity and battery costs, battery-electric locomotive operating costs are only lower than diesel-electric locomotive operating costs if the environmental impacts are factored into the economic calculation and under the assumption that typical 241km range can be achieved by single boxcar with 14MWh battery (*14*).

## 3 METHODOLOGY

This framework simulates and evaluates the deployment of alternative fuel technologies on the rail network, and as such is dependent on the technological requirements of each energy technology. The energy technologies considered in this paper can be divided into two categories: (1) lower-carbon drop-in fuels such as biodiesels and e-fuels and (2) new energy storage technologies such as battery-electric or hydrogen, as discussed in Section 1.

Lower-carbon drop-in fuels are generally deployed as admixtures (e.g., 20% biodiesel and 80% diesel). Their deployment is considered uniform across the network at a desired penetration rate taken as the rate of the admixture. Thus, origin-destination (O-D) flows are simply routed by commodity group on the existing (baseline) network's links. These flows are used to calculate the costs and emissions associated with their deployment by weighing the relevant fuel parameters with their corresponding admixture rates. Figure 1 demonstrates the essential steps and flow of information.

However, for alternative energy sources that require locomotive powertrains to be converted and new refueling infrastructure to be deployed, we present a five-step sequential framework to (1) select O-D paths that leverage economies of density, (2) locate facilities along these paths, (3) reroute flows on the existing and alternative technology rail networks, (4) size the facilities in order to serve the rerouted flows, and (5) evaluate the deployment in terms of their emissions, costs, and operational impacts. The flow of information between each of the steps can be seen in Figure 2.

### 3.1 Sequential Framework for Facility Location & Sizing

The joint facility location and sizing problem is a combinatorial problem to solve (*47*), especially over networks, where potential facility locations have many degrees of interconnectivity. To simplify the problem, therefore, we decouple and formulate variations of the facility location and sizing problems that capture important managerial concerns.

We formulate the facility location integer program as an extension of the Set Covering Problem (*48*). The solution yields the minimum number of facilities on the network required to fulfill continuous trips along the paths between a pre-specified set of O-D pairs. These O-D pairs are selected from the network based on the flows between them. Each O-D pair is ranked in descending order by the value of the ton-miles of goods moved between them, from which a subset is selected based on the input O-D coverage ratio (see Figure 2). Paths are generated between each O-D pair in this subset and are used, along with the input locomotive range, to specify the coverage constraints for the integer program. To locate facilities, the existing rail network is represented as a directed graph in which the nodes represent candidate locations—rail yards with the capacity to service trains—for refueling/charging facilities and the arcs represent the railroad tracks between the candidate facility locations. The selected facilities are used to create a subnetwork enabled by the alternative technology. To represent where flows can be served by the alternative technology, the enabled arcs on this subnetwork are either (1) along paths between facilities with a distance below the maximum given range or (2) on paths with a distance of no more than half the given range (i.e., the in-and-back-out distance from a facility does not exceed the range).

As flows are assumed to originate and terminate at any candidate facility location on the rail network, they may be assigned to the alternative technology enabled subnetwork. The actual assignment of flows depends on the specified routing policy, which may or may not allow for flows to be rerouted from their original routing on the baseline network, may have a maximum distance increase if flows are rerouted, or may only allow flows originating *and* terminating at enabled nodes to be served by the alternative technology. These, and other routing strategies can be accommodated, as specified by the input to the Flow Routing step. Importantly, as the alternative technology may not be able to serve all flows on the network, the baseline network, operated by diesel, is assigned any flows that cannot be served by the alternative technology. We assume that for a given O-D pair, the flows are served entirely by the alternative technology (if service is enabled), or entirely by diesel (as in the baseline case). The routing of flows on the two networks is used to compute the penetration rate (in percentage of ton-miles) of the alternative technology on the network.

As the energy intensities of moving different commodities are known to vary considerably (*49*), for each O-D pair, we route the flows between them for each of the nine main commodity groups as reported by the AAR (*50*). The selected facility locations and commodity-specific link-flows for the alternative technology subnetwork are critical inputs for the sizing of facilities (i.e., the specification of facility energy capacities). We build on the minimum cost network flow problem structure (*51*) to formulate a facility sizing problem which provides the energy flows at each of the refueling/charging facilities that minimizes the total cost of energy consumption for the network. The constraints for this formulation ensure the energy required to move all goods (calculated by commodity) is dispensed by the selected facilities. Peak link-wise energy requirements are calculated as the product of the commodity-specific peak flows assigned to each link and the commodity-specific energy intensities. The model outputs the peak facility size (in kWh/day for battery and $kgH_2$/day for hydrogen) of each selected facility that is required to provide service to the goods routed on the alternative technology's network as well as the average energy consumption (kWh/day or $kgH_2$/day) and locomotive throughput (locomotive/day). These outputs are critical to calculating each facility's utilization rate as well as the emissions, cost, and operational impacts of a particular deployment.

## 3.2  Life-cycle Analysis (LCA) of Energy Technologies

In this study, we examine the GHG emissions of different energy technologies with a system boundary covering both the well-to-pump (WTP) and pump-to-wheel (PTW) stages. The functional unit for the emissions was set as $gCO_2$/ton-mile. Together, WTP and PTW stages comprise well-to-wheel (WTW) analysis. While the WTP examines the environmental impact of production, transportation, and distribution of feedstock and fuels, PTW focuses on the vehicle operation. Note that the impacts from the vehicle manufacturing cycle, including stages such as material extraction, component manufacturing, assembly, and recycling of vehicle components, are out of scope for this analysis.

We use ANL's GREET model (*15*)—updated annually with the most up-to-date and detailed energy use and emissions data for petroleum refineries and electric power plants—to conduct the WTW analysis. For the diesel, biodiesel, e-fuel, and hydrogen (for various fuel pathways, e.g., steam-methane reformed or renewable hydrogen) technologies, we estimate the WTW GHG emissions factors (in gCO$_2$/Btu) using GREET 2021 (*15*). The R-1 report published by Surface Transportation Board (STB) provides the annual diesel usage and associated revenue ton-miles for each of the Class I railroads (*52*). Combining these values with the emissions factors from GREET, we estimate the railroad-specific WTW GHG emissions in gCO$_2$/ton-mile using Equation (1).

$$GHG\ emissions\ \left[\frac{gram\ CO_2}{ton-mile}\right] = \frac{Total\ Diesel\ Use\ [gallons]}{Total\ Revenue\ Ton-Miles\ [ton-mile]}$$
$$\times\ Emission\ Factor\ from\ GREET\ \left[\frac{gram\ CO_2}{Btu}\right] \quad (1)$$
$$\times\ Lower\ Heating\ Value\ of\ Diesel\ \left[\frac{Btu}{gallons}\right]$$

For battery-electric locomotives, we consider the GHG emissions associated with the upstream processes of electricity generation and the losses in the transmission and distribution system. For current and projected (Mid-case Standard Scenario) generation mixes, we capture the state-wise variation of electricity generation mixes in terms of gCO$_2$/kWh supplied to the charger station based on the results from (*53*).

### 3.3  Techno-economic Analysis (TEA)

For conventional diesel, biodiesel, and e-fuels, this study focuses on the levelized cost of refueling only, as these energy technologies do not require additional infrastructure investments. We estimate the levelized cost of refueling by multiplying the commodity specific link flow, energy intensity, and fuel cost. For battery-electric and hydrogen technology, we consider the charging/refueling infrastructure cost in addition to the battery/hydrogen tender car capital investment and refueling costs. The sequential framework described in Section 3.1 estimates the facility size based on the peak demand (in kWh/day for battery and kgH$_2$/day for hydrogen) at each location. Based on a given locomotive battery/hydrogen tender car capacity, the peak and average demand is used to estimate the peak and average locomotive throughput at each facility. From the peak locomotive throughput, we estimate the number of chargers/pumps needed at each location to support the peak demand, such that a provided maximum station utilization is not exceeded. ANL's bottom-up TEA tools, Heavy-duty Electric Vehicle Infrastructure Scenario Analysis Model (HEVISAM) for battery-electric, and Hydrogen Delivery Scenario Analysis Model (HDSAM) for hydrogen (*54*), provide the levelized cost of charging/refueling for a given fleet size and facility specification. We use HEVISAM to develop a functional relationship between the levelized cost of charging and number of locomotives for a given number of chargers. The levelized cost of operation for the battery-electric technology is calculated using Equation (2), while the levelized cost of operation for the hydrogen technology scenarios is represented in Equation (3).

$$\begin{aligned}
\text{Levelized cost of operation} &\left[\frac{\$}{kWh}\right] \\
&= \text{levelized cost of battery (ammortized capital cost of battery)} \\
&+ \text{levelized cost of charging (charging station contribution from HEVISAM)} \\
&+ \text{electricity price by state}
\end{aligned} \quad (2)$$

$$\begin{aligned}
\text{Levelized cost of operation} &\left[\frac{\$}{kgH_2}\right] \\
&= \text{levelized cost of hydrogen tender car (ammortized capital cost of tender car)} \\
&+ \text{levelized cost of refueling (delivery and refueling station capital cost contribution from HDSAM)} \\
&+ \text{hydrogen fuel price (production)}
\end{aligned} \quad (3)$$

The levelized costs of operation are estimated in terms of cost per quantity of energy (e.g., $/kWh) or fuel (e.g., $/kgH$_2$, $/gallon) and are converted to cost per revenue ton-mile using energy intensity parameters from Table 1 and the results from the flow routing and facility sizing steps to determine the ton-miles served by a particular energy technology. The levelized cost per ton-mile provides a standardized way to represent alternative technology costs as an operational metric.

The WTW GHG emissions and the levelized cost of operation could be synthesized into one metric to compare across energy technologies. The cost of avoided emissions (CAE) of a particular technology is the ratio of the levelized cost of operations (in $/ton-mile) and the WTW GHG emissions intensity (in kgCO$_2$/ton-mile), relative to the baseline diesel operations, as represented in Equation (**4**). The CAE serves as a key policy metric, as it provides the cost per unit of carbon reduced from emissions for a specific technology, a metric that can be compared to the social cost of carbon or carbon credit/tax schemes.

$$\text{Cost of avoided emissions}\left[\frac{\$}{kgCO_2}\right] = \frac{\text{LCO alternative technology} - \text{LCO diesel}}{\text{WTW GHG alternative technology} - \text{WTW GHG diesel}} \quad (4)$$

## 4  APPLICATION DATA SOURCES AND PARAMETERS

The framework presented above is next illustrated through application to evaluate alternative decarbonization scenarios for the US Class I railroads network. In this application, the parameters involved in modeling rail operations and energy sources collected from multiple sources are summarized in Table 1.

**Table 1 - Parameters**

| | **A. Train Operations** | | |
|---|---|---|---|
| **Parameter** | **Western Railroads** | **Eastern Railroads** | **Source** |
| Freight demand | *(various, by O-D by commodity)* | | *(55)* |
| Average number of locomotives per train | 3.15 | 2.18 | *(52)* |
| Average number of cars per train | 74.6 | 68.5 | *(52)* |
| Average tonnage per locomotive | 1319 | 1403 | *(52)* |

| Marginal battery cost per locomotive (¢/ton-mile) | 0.12 | 0.19 | *(14, 52)* |
|---|---|---|---|
| Marginal hydrogen tender car cost per locomotive (¢/ton-mile) | 0.05 | 0.08 | *(52)* |
| Average hydrogen locomotive range (mile) | 1039 | 977 | *(15, 52)* |
| Diesel Energy Requirement for Various Commodities (BTU/ton-mile) | | | |
| Agricultural & Foods | 152 | 155 | *(15, 43, 50, 52, 56)* |
| Chemical & Petroleum | 150 | 153 | |
| Coal | 107 | 109 | |
| Forest Products | 219 | 224 | |
| Intermodal | 875 | 893 | |
| Metals and Ores | 152 | 155 | |
| Motor Vehicles | 710 | 725 | |
| Nonmetallic Products | 128 | 131 | |
| Others | 553 | 565 | |
| **B. Battery-Electric** | | | |
| Parameter | Value | | Source |
| Unit weight of battery tender car (ton) | 150 | | *(14)* |
| Battery capacity (MWh) | 14 | | *(14)* |
| Charging speed (MW) | 3 | | *(14)* |
| Charging depth | 80% | | *(14)* |
| Battery energy efficiency | 95% | | *(14)* |
| Capital cost of battery + inverter + boxcar ($) | 1,271,816 | | *(14)* |
| Future cost of battery ($) | 452,908 | | *(14)* |
| Battery maintenance cost ($/day) | 100 | | *(14)* |
| Battery lifetime (year) | 13 | | *(14)* |
| Relative energy efficiency of battery-electric to diesel | 2.44 | | *(14)* |
| Discount rate | 3% | | *(14)* |
| Time horizon (year) | 26 | | *(14)* |
| Charging cost ($/kWh) | 0.15 | | *(15)* |
| Electric grid carbon emissions (kg $CO_2$ eqv/kWh) | (various, by state by year) | | *(53)* |
| Electric grid cost ($/kWh) | (various, by state by year) | | *(57)* |
| **C. Hydrogen** | | | |
| Tender car capacity (kg$H_2$) | 4000 | | *Assumed* |
| Tender car capital cost ($/kg$H_2$) | 80 | | *Assumed* |
| Tender car lifetime (year) | 20 | | *Assumed* |

| Relative energy efficiency of hydrogen to diesel | 1.5 | *(58)* |
|---|---|---|
| Hydrogen emissions (kg $CO_2$ eqv/kg$H_2$) | 14.77 | *(15)* |
| Hydrogen fuel cost ($/kg$H_2$) | 2.00 | *(54)* |
| **D. Drop-in Fuels** | | |
| **Parameter** | **Value** | **Source** |
| Diesel lower heating value (BTU/gallon) | 129,488 | *(15)* |
| Relative energy efficiency of drop-in fuels to diesel | 1 | *(15)* |
| Diesel emissions (kg $CO_2$ eqv/gallon) | 12.36 | *(15)* |
| Biodiesel emissions (kg $CO_2$ eqv/gallon) | 3.50 | *(15)* |
| E-fuel emissions (kg $CO_2$ eqv/gallon) | 0.07 | *(15)* |
| Diesel cost ($/gallon) | 2.47 | *(59)* |
| Biodiesel cost ($/gallon) | 3.60 | *(59)* |
| E-fuel cost ($/gallon) | 5.19 | *(59)* |

## 4.1 Rail Network

The existing rail network was extracted from the North American Rail Network data set compiled by the Federal Railroad Administration (FRA) and the Bureau of Transportation Statistics (BTS) using work on rail facility classification done by Oak Ridge National Lab in WebTRAGIS (*61*). These facility classifications allow for the number of nodes on the network to be reduced from the tens of thousands to hundreds as those nodes representing terminal, primary, and minor rail yards are kept and all others—representing grade crossings—were removed. Additionally, nearby nodes are clustered into super-nodes to simplify the network topology, while maintaining its overall structure. Operational data from the Annual Report of Finances and Operations (R-1 report) (*52*) on values for average train loadings, annual movements of goods and fuel consumption, locomotive fleet sizes, and physical train parameters were extracted for each of the Class I railroads. The relevant values that were calculated from these data are summarized in Table 1A.

Freight rail demand for 2019 was estimated from the STB's annually compiled Carload Waybill Sample (CWS) (*55*), which samples a subset of all rail movements in the U.S. and provides movement-specific data on railroad, routing, and costs. Though this framework can be applied to any individual railroad, the CWS data was aggregated to the three-railroad level in accordance with STB policy to preserve confidentiality in the illustration of results that follow. All operational parameters were also aggregated in a similar manner. As the aggregated CWS provides O-D flow data, these must be routed on the network following an assumed routing policy.

### 4.1.1 Energy requirement

As the flow routing from the CWS provides estimates of the ton-miles of goods by commodity on the network, these must be converted into terms of energy from which fuel or electricity consumption can be calculated. The energy required to move a ton of goods one mile varies by commodity due to differences in physical characteristics (e.g., density) and operational practices (e.g., shipment speed), by railroad due to topographical and fleet variations, and by energy technology due to differences in powertrain designs (*49*). As such, various factors must be applied to each ton-mile of goods based on its commodity type, the railroad that is moving it, and the locomotive's energy technology. A tool for calculating commodity-specific energy intensity factors was developed in (*56*), with the values updated for 2019 freight rail operations in (*15*). Operational data from the R-1 report was used to regroup these commodity-specific energy intensity factors into the nine main commodity groupings recorded by the AAR (*50*). Additionally, data on railroad-specific fuel consumption and ton-mile service from (*52*) was used to calculate railroad-specific energy intensities in (*43*), which were then reaggregated to the corresponding groupings used in the scenarios to follow. Finally, technology-specific energy efficiency ratios were taken from various sources. For biodiesel and e-fuels we assume the same energy intensity as for diesel. For battery-electric locomotives, the efficiency gain from battery was estimated as 2.44 the quotient of the battery round trip efficiency (assumed to be 95%) and diesel engine efficiency (assumed to be 39%) (*14*). For Hydrogen locomotives, the energy efficiency gain compared to baseline diesel was taken as 1.5 (*58*). These factors are shown in Table 1A.

### 4.2 New Energy Sources

New energy sources are evaluated relative to diesel operations based primarily on their differences in cost and emissions. Baseline diesel cost and emissions data are extracted from the R-1 report (*52*). Forecasts for drop-in fuel costs and emissions are based on (*59*) and are summarized in Table 1D. For battery-electric locomotives, the electric grid has geographically varying costs and emissions, which directly affect the evaluation of battery-electric deployment. State-specific commercial electricity rates from (*57*) and emissions values from (*53*) were used. In addition to electrical costs, the economic evaluation of the battery-electric scenario must consider the levelized capital cost of charging facility deployment. Thus, the TEA tool, HEVISAM, is applied using data found in Table 1B attained from prior work and personal communication with industry experts to estimate the levelized cost of operation for a given charging facility depending on its capacity and utilization rate. For hydrogen fuel locomotives, emissions values from (*15*) are used to estimate the WTW emissions associated with the use of steam-methane reformed hydrogen fuel, while refueling station capital costs and hydrogen fuel costs are provided by the HDRSAM tool using the parameters in Table 1C.

#### 4.2.1 Levelized Cost of Battery

For the case of battery-electric locomotive deployment, the sizeable capital cost of the batteries must be captured. Building on assumptions and data from (*14*) on the cost of a 14 MWh battery tender car attachment for a locomotive, we calculate the levelized cost of battery tender car operation in dollars per ton-mile based on operational data from (*52*). These values vary by railroad as seen in Table 1A and are a component of the complete levelized cost in Equation (2). The framework takes locomotive range as an input that is used to calculate the energy storage capacity assigned to each locomotive (in the form of additional battery tender cars) based on the average tonnage assigned to each locomotive. As the locomotive range increases, so does the required energy storage per locomotive, which in turn increases the energy capacity of battery tender cars per locomotive and levelized cost of battery operations.

*4.2.2   Levelized Cost of Hydrogen Tender Car*

Due to considerably lower energy density relative to diesel, hydrogen fuel locomotives will require the use of a tender car for fuel storage. The storage of hydrogen fuel requires advanced temperature and pressure control systems, making hydrogen tender cars capital-intensive investments (*62*). Using the techno-economic data summarized in Table 1C, we amortize the capital cost of the hydrogen tender car over its lifetime. Operational data from the R-1 report in (*52*) is then used to estimate the cost per ton-mile of hydrogen tender car operation, which vary by railroad as seen in Table 1A. These values are factored into the complete levelized cost of operation in Equation (3). From the assumption on fixed liquid hydrogen tender car capacities at 4000 $kgH_2$, we are able to estimate hydrogen locomotive range based on each railroad's average locomotive payload, as seen in Table 1A.

# 5   RESULT ANALYSIS

To illustrate the functionality of the model, scenario simulation and evaluation results are shown for Western (BNSF, Canadian National, and Canadian Pacific Railways) and Eastern (CSX Transportation, Kansas City Southern, and Norfolk Southern Railways) railroad networks.

Figure 3 shows an example of the NUFRIEND dashboard. It allows users to model different scenarios based on inputs including railroads, energy sources, commodity groups, battery ranges, and target deployment percentages on the left, as well as other specific parameters on a separate pane. In the context of battery-electric and hydrogen deployment, the five-step sequential framework is applied to consider the railroad network and freight traffic, locate and size the charging facilities, route the rail traffic, and estimate the emissions and costs based on LCA and TEA, as outputted in the right. Metrics including WTW emissions, levelized costs of operation (LCO), the proportion of ton-miles served by each energy technology, the cost of avoided emissions, and other detailed operational results are shown. The scenario WTW emissions are the sum of emissions of diesel (blue) and battery/hydrogen (green) routes. The LCO of diesel is the current fuel cost, while the battery-electric LCO is composed of charging facility capital costs, battery capital and O&M cost, and electricity cost and the hydrogen LCO is composed of refueling facility capital costs, energy tender car capital cost, and fuel cost. Above the battery-electric/hydrogen LCO, the scenario average LCO is shown, which includes the cost of diesel refueling needed to serve the segments of the network not covered by battery-electric or hydrogen. Results are also shown down to the facility and track levels with green color denoting coverage of battery-electric or hydrogen technology. Users can hover over them to examine more granular information such as traffic volume, number of chargers or pumps, and station utilization.

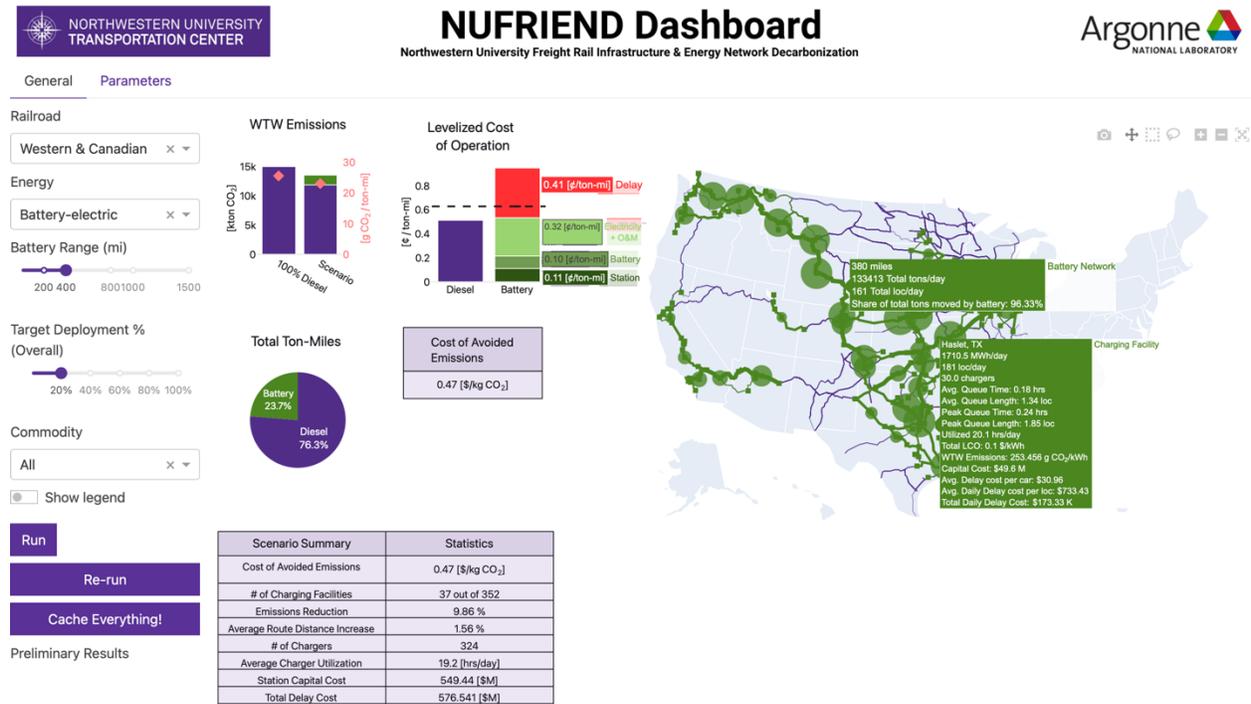

Figure 3 - NUFRIEND Dashboard Example

## 5.1 Example Illustration

This section showcases the functionalities of the dashboard with scenario results for each of the two energy technology categories. Each of the figures show aggregate emissions, cost, and deployment plots, as well as a network map that shows the specific alternative technology deployment. The WTW emissions bar plots show the emissions (in kton $CO_2$) attributable to diesel (blue) and the alternative technology (green) operations, with the emissions intensity (in g $CO_2$/ton-mile) of diesel and the deployment scenario in question overlaid (yellow diamond). The levelized cost of operation bar plots show the LCO (in ¢/ton-mile) for diesel and the alternative technology in question. The key cost components are displayed, such as fuel (blue) for diesel, biofuel, e-fuel, and hydrogen; electricity (light blue) for battery; battery and hydrogen fuel tender car costs (orange); and charging/refueling station capital costs (red). The pie chart shows the deployment of the alternative technology as the share of ton-miles captured by diesel (blue) and by the alternative technology (green).

### 5.1.1 Battery-Electric and Hydrogen

Figure 4 and Figure 5 show an example (hypothetical) deployment of a 400-mile range battery-electric locomotive technology for Western and Eastern railroads, respectively, needed to serve approximately 50% of their ton-mileage. Western rail networks are in general more expansive and require more charging facilities compared to Eastern railroads (57 vs 21). This also leads to difficulties in connecting the whole network and affects the overall cost of avoided emissions. The emissions and costs associated with battery-electric technologies are highly dependent on those of the electric grid and therefore sensitive towards the locations of the charging facilities. To reduce emissions by one kilogram of $CO_2$, it costs Western railroads $0.11 and Eastern railroads $0.09 under the examined deployment scenario.

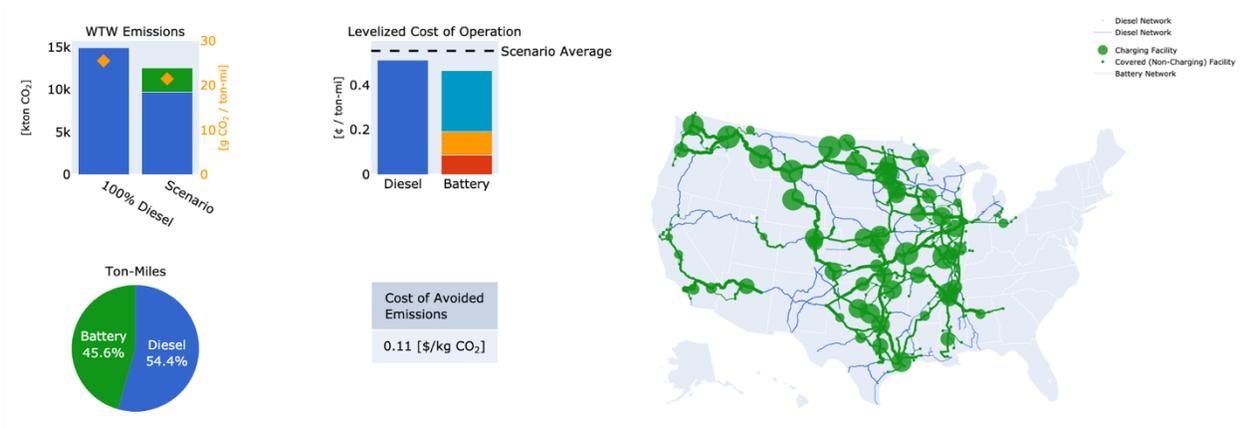

**Figure 4 – Example of Battery-Electric Deployment for Western Railroads**

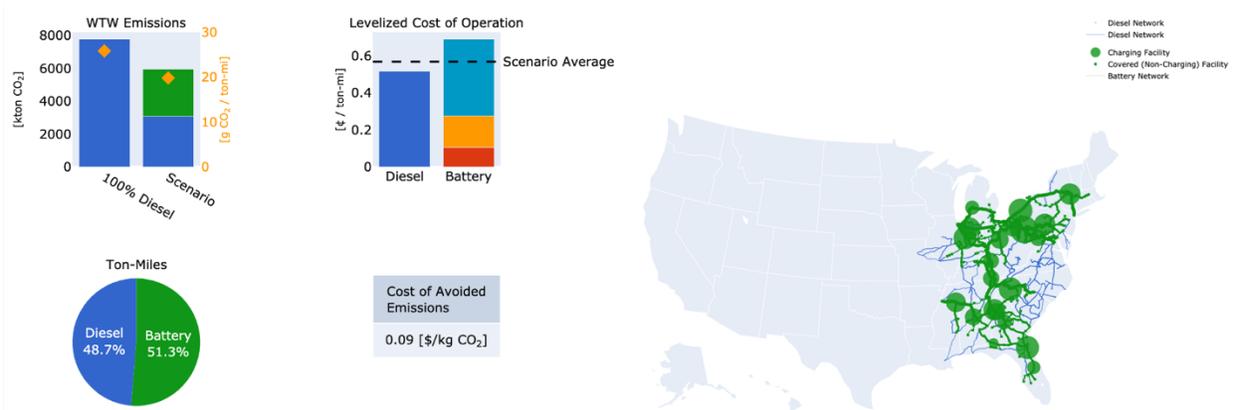

**Figure 5 – Example of Battery-Electric Deployment for Eastern Railroads**

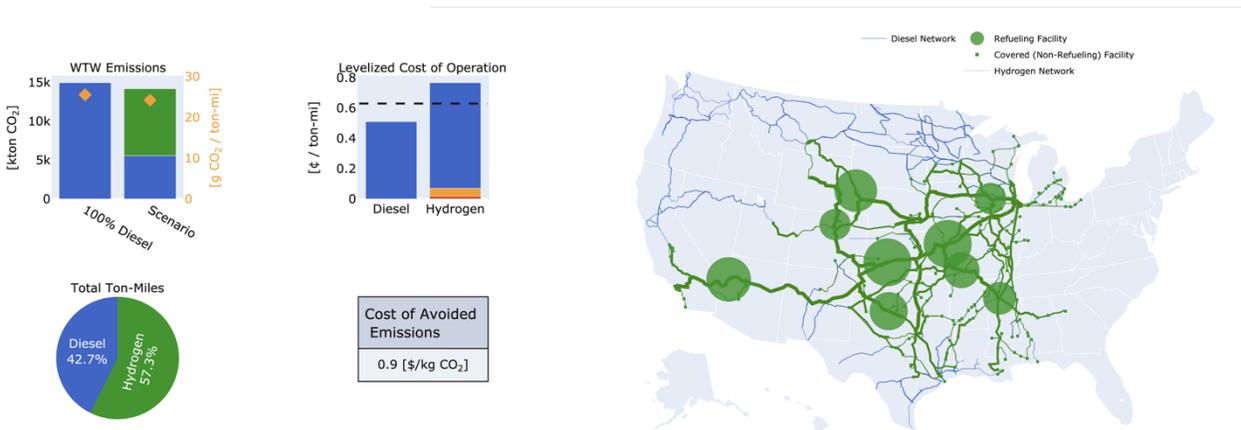

Figure 6 and Figure 7 show a corresponding example deployment of a hydrogen locomotive technology (with an approximate 1000-mile range) for Western and Eastern railroads, respectively, needed to serve approximately 55% of their ton-mileage. Though Western rail networks are in general more expansive than Eastern rail networks, the long range of this hydrogen locomotive diminishes the number of required refueling facilities compared to the battery case (9 for Western vs 3 for Eastern). Most notably, hydrogen locomotive operations are not considerably cleaner than diesel operations, as hydrogen fuel is primarily produced through natural gas reforming. Furthermore, liquid hydrogen fuel exhibits high costs, due to the energy intensive process of onsite liquefaction. The relatively low emissions reductions and high incremental costs lead to high costs of avoided emissions. To reduce emissions by one kilogram of $CO_2$ through hydrogen operations, it costs Western railroads $0.9 and Eastern railroads $1.05 under the examined deployment scenario. More environmentally friendly (e.g., solar, renewable, nuclear powered) and economical processes for hydrogen production would be required to help hydrogen decarbonize the rail freight sector.

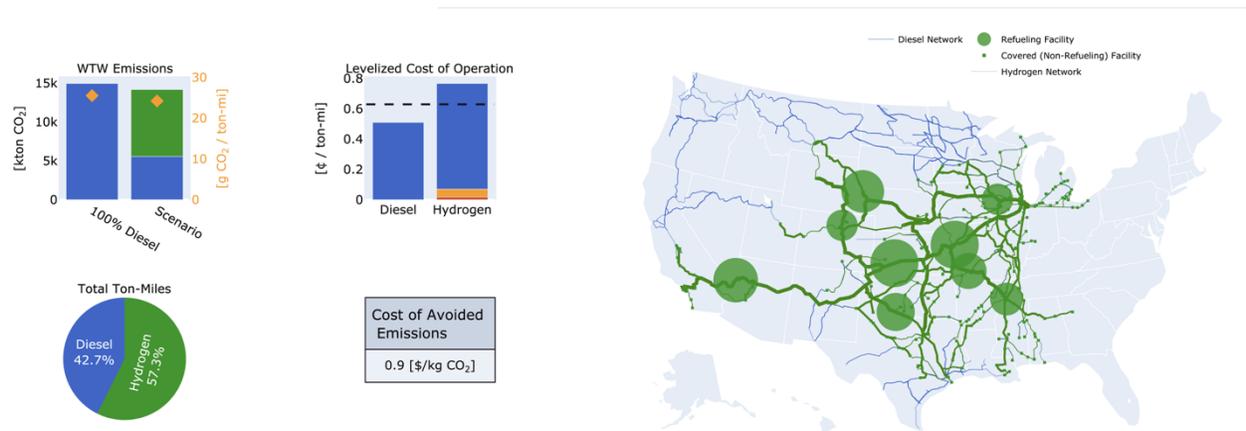

**Figure 6 – Example of Hydrogen Deployment for Western Railroads**

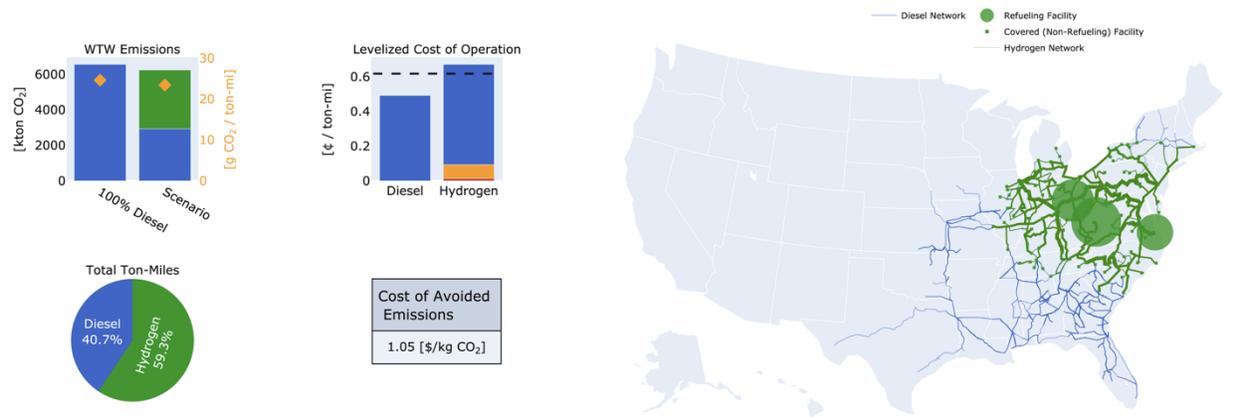

**Figure 7 – Example of Hydrogen Deployment for Eastern Railroads**

*5.1.2   Biofuel and E-fuel*

For drop-in fuels such as biofuels and e-fuels, fuel blends are assumed to be applied uniformly across all locomotives on the network. Figure 8 and Figure 9 show the results for 50% deployment of biofuels in Western and Eastern railroads, respectively. This deployment of biodiesels would contribute to a 36% reduction in emissions (for both railroad groups) relative to diesel, with a cost of $0.13 per kilogram of CO2 reduced. Figure 10 and Figure 11 show the results for 50% deployment of e-fuels in Western and Eastern railroads, respectively. As e-fuels are nearly carbon neutral, they would provide a more promising environmental solution than biofuels, albeit at significantly greater cost (nearly double that of conventional diesel). Their deployment in this scenario would contribute to a 50% decrease in carbon emissions at a cost of $0.22 per kilogram of CO2 eliminated.

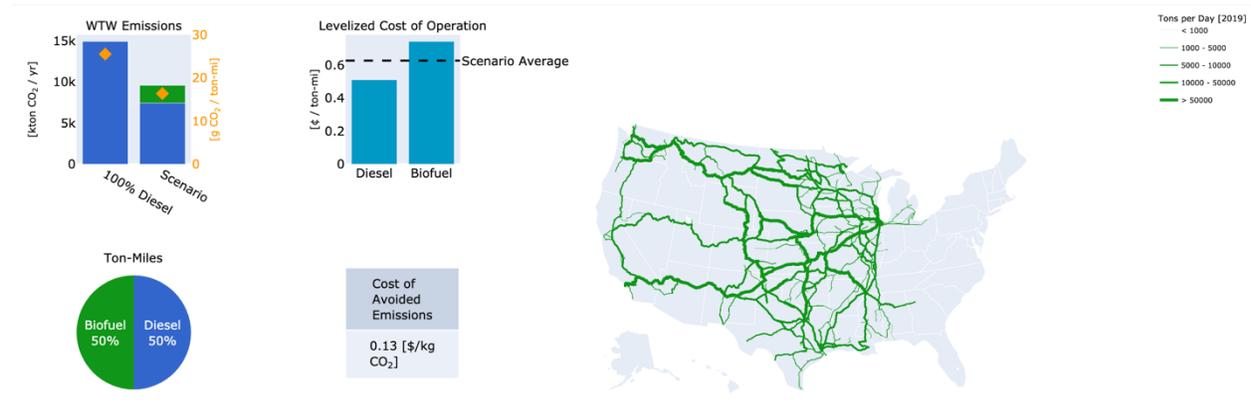

**Figure 8 – Example of 50% Biodiesel Deployment for Western Railroads**

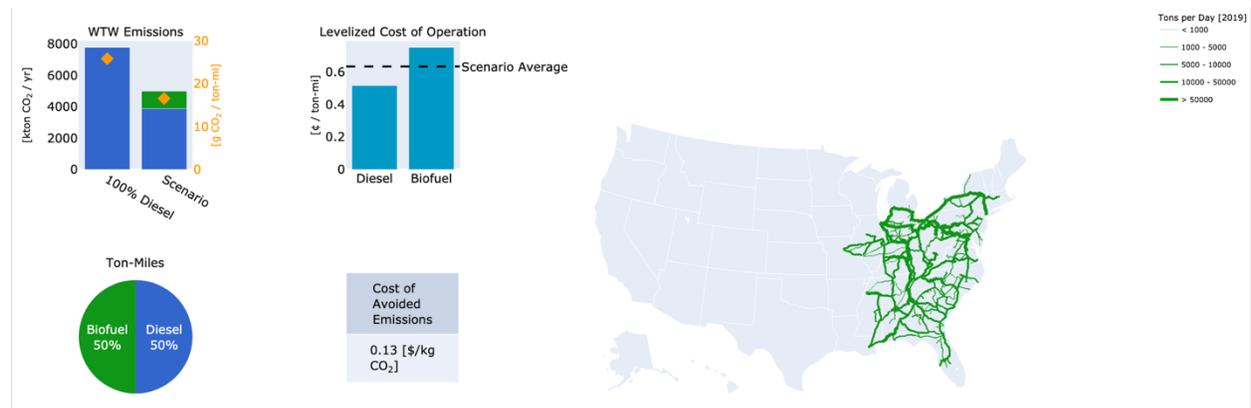

**Figure 9 – Example of 50% Biodiesel Deployment for Eastern Railroads**

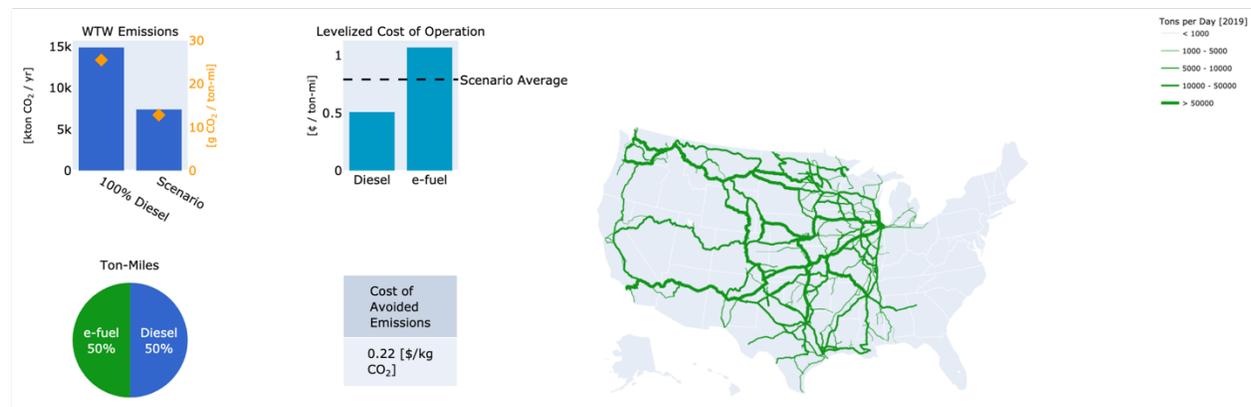

**Figure 10 – Example of 50% E-fuel Deployment for Western Railroads**

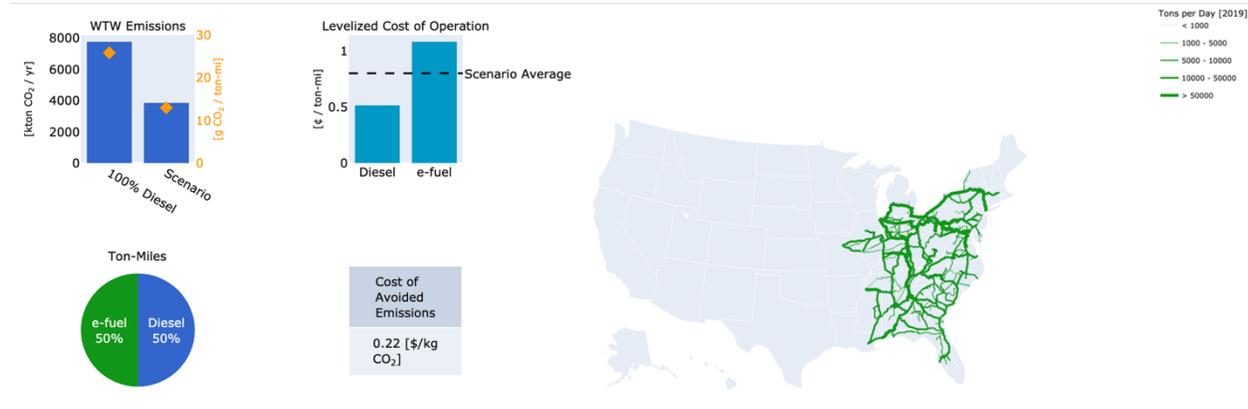

**Figure 11 – Example of 50% E-fuel Deployment for Eastern Railroads**

## 5.2 Scenario Comparison

This section evaluates the optimization and simulation results of the battery-electric deployments, demonstrating the framework's potential in analyzing and comparing across scenarios.

### 5.2.1 Deployment Percentage

A key functionality of our framework implemented in the dashboard allows users to input different deployment percentages to simulate, optimize, and evaluate different intermediate stages of a technology's roll-out.

Figure 12-Figure 14 show the results for Eastern railroads under (approximately) 30%, 50%, and 100% target deployments. As the deployment percentages increase, the network coverage increases drastically, with the number of facilities growing from 12, through 21 to 167 at 100%. This highlights the high number of facilities required to serve the "last miles" of the rail network. While emissions decrease rather proportionally with the extent of the roll-out, the initial LCO at 30% deployment is considerably high, due to starting costs on capital infrastructure projects. Higher deployment percentages enable economies of scale to reduce the costs of avoided emissions from $0.17/kg $CO_2$ at 30% to $0.09/kg $CO_2$ at 50%.

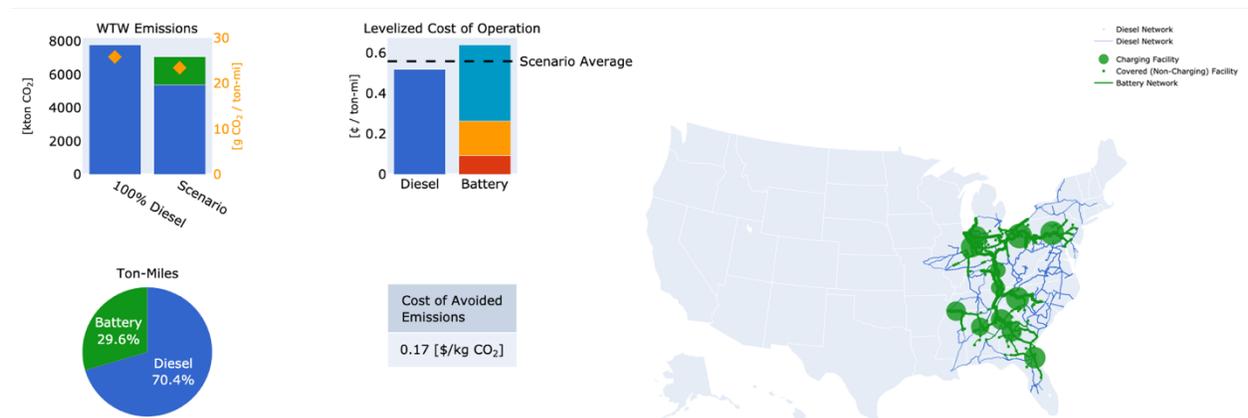

**Figure 12 – Results for Eastern Railroads with 30% Target Deployment Percentage**

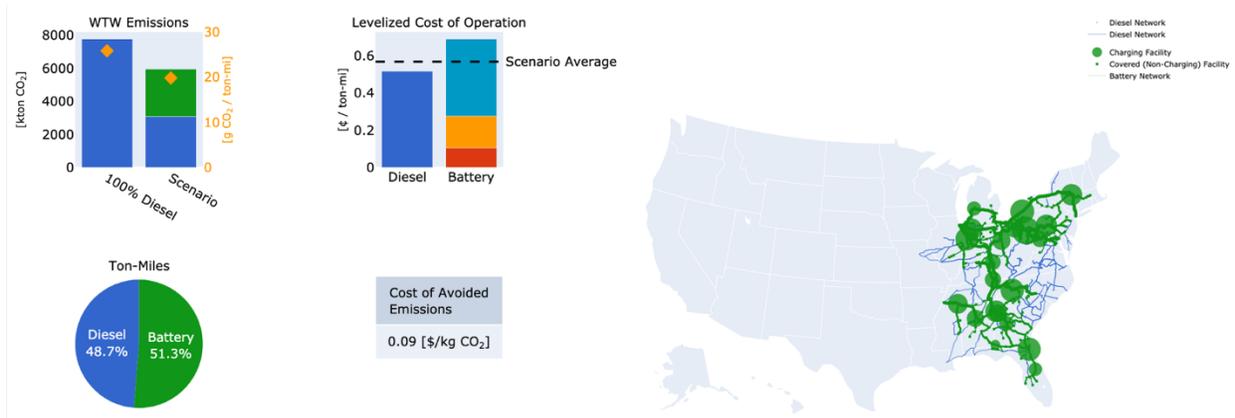

**Figure 13 – Results for Eastern Railroads with 50% Target Deployment Percentage**

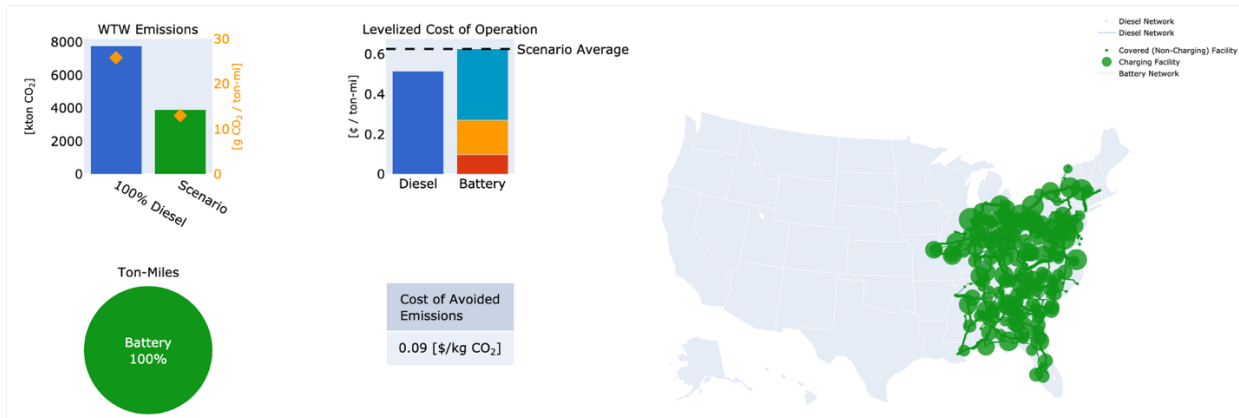

**Figure 14 – Results for Eastern Railroads with 100% Target Deployment Percentage**

*5.2.2  Range*

In the context of battery-electric deployment, locomotive range affects both the economics and environmental performance of a particular scenario.

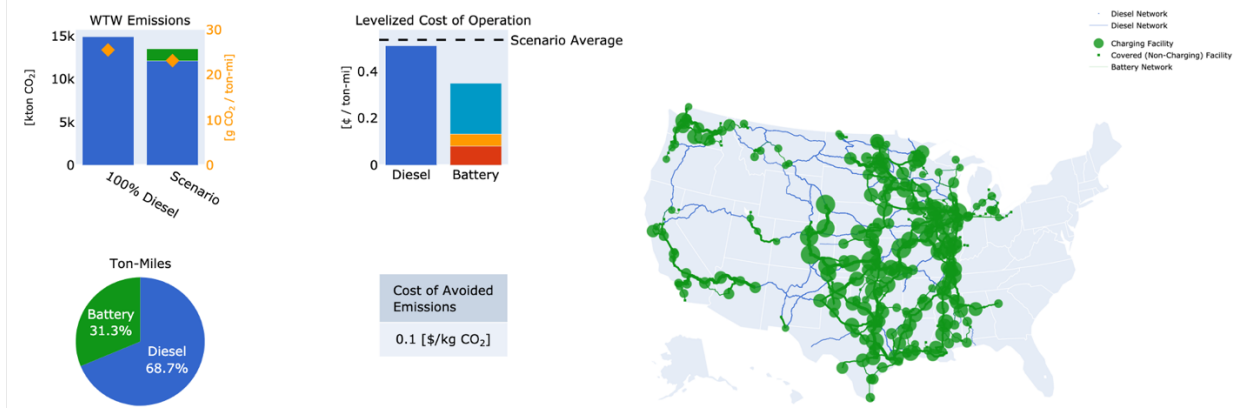

Figure 15-Figure 17 illustrate the key trade-off associated with locomotive range. Increasing the locomotive range increases the energy storage capacity per locomotive and thus increases the total battery purchase and operating costs. However, an increase in locomotive range allows for greater

reach and fewer charging stations to be deployed, reducing the total facility capital costs, while increasing the network penetration of service that can be provided. This trade-off is illustrated by the costs of avoided emissions which are \$0.10/kg $CO_2$ at 200-mile range, \$0.11/kg $CO_2$ at 400-mile range, and \$0.06/kg $CO_2$ at 800-mile range. The stark decrease in the cost of avoided emissions between the 400-mile and 800-mile range cases comes from the consolidation of freight along key corridors that allows emissions reductions to go from 16% with 400-mile range locomotives to 46% with 800-mile range locomotives. Furthermore, locomotives with longer ranges (e.g., 800 miles) can significantly reduce emissions, as they can be used to decarbonize more energy intensive commodities (i.e., intermodal) that are typically shipped over long distances. Note that with locomotives with a 200-mile range, though a 50% target deployment level was set, only 31% of ton-miles could be served by battery-electric locomotives, due to the insufficient range on the expansive Western railroad network. The flexibility of the range parameter supports the sensitivity analysis of optimal technology range values.

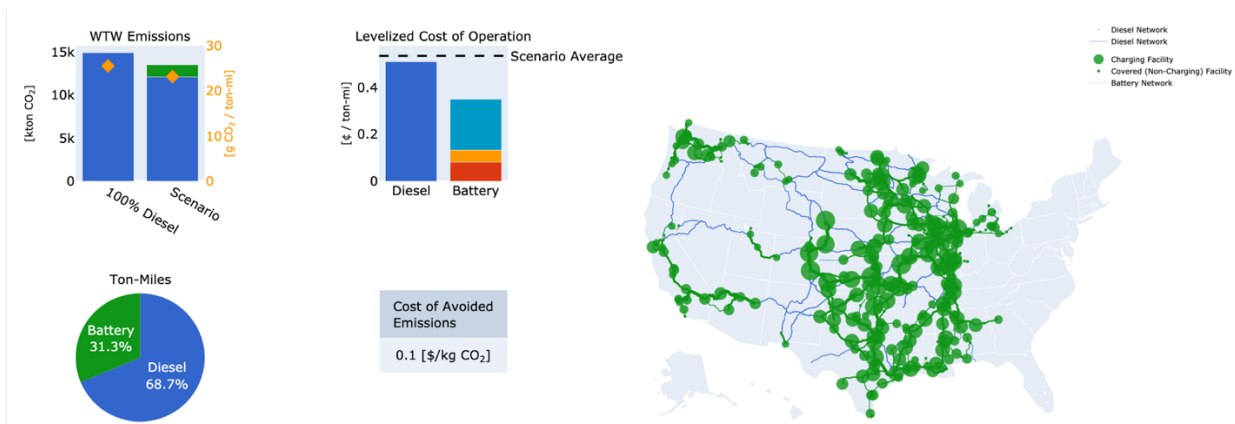

**Figure 15 – Results for Western Railroads with 200-mile Range**

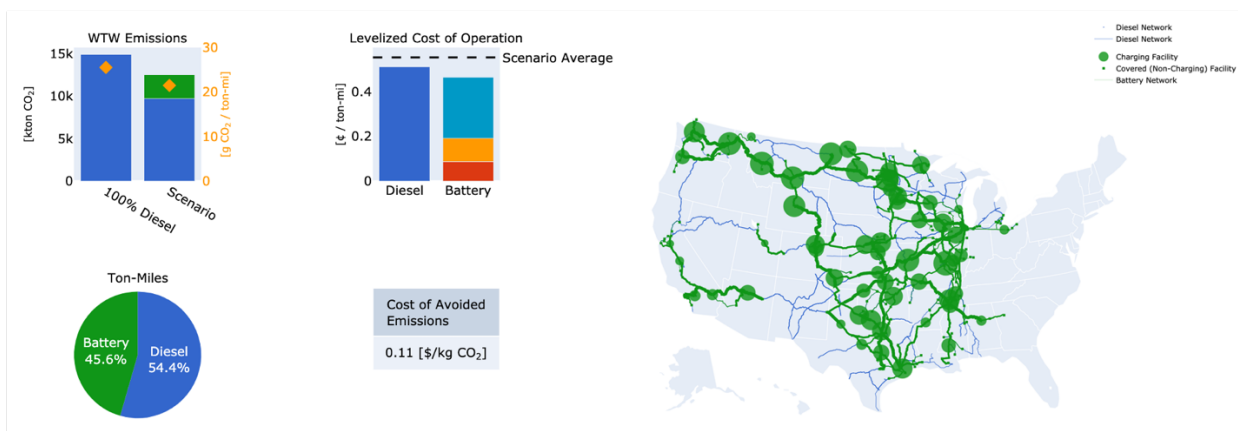

**Figure 16 – Results for Western Railroads with 400-mile Range**

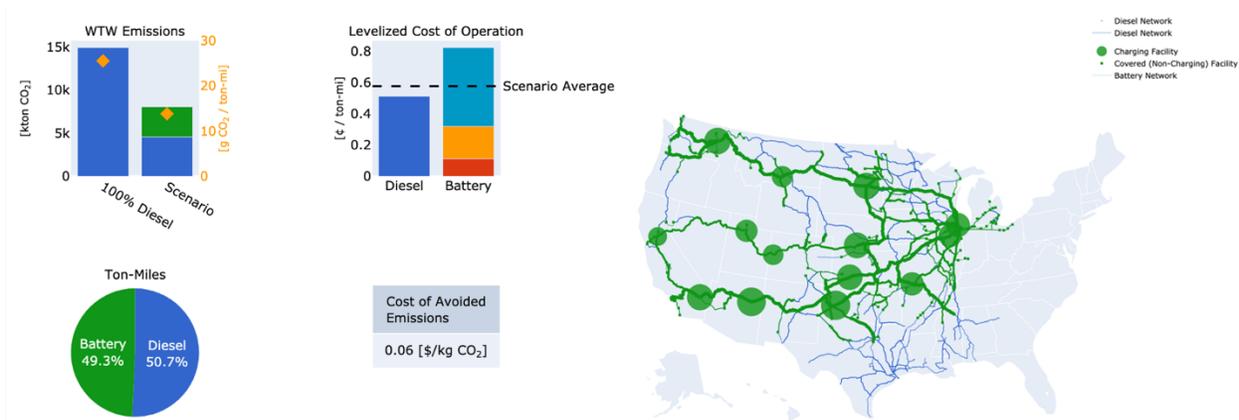

**Figure 17 – Results for Western Railroads with 800-mile Range**

## 6 CONCLUSION

This paper presents a simulation framework to aid stakeholders optimize, plan, and analyze deployment strategies for alternative fuel propulsion technologies in freight rail. It is applied to evaluate different classes of energy technologies that can be used to decarbonize freight rail such as biofuels, e-fuels, hydrogen, and battery-electric pathways. The five-step approach developed in this work builds on nominal graph theory problems to simplify and solve otherwise intractable problems in the field of facility location and sizing. This framework can efficiently simulate a variety of technology adoption scenarios with built-in LCA and TEA tools and features input flexibility for evaluating a variety of deployment scenarios and performing sensitivity analysis of key technological and operational parameters. Key metrics covering emissions, costs, facility locations and sizes, and traffic volumes are outputted in aggregate and granular levels for detailed analysis.

The framework's functionalities are demonstrated with US Class I railroad data for Eastern and Western railroad networks and technological parameters for battery-electric, biofuel, and e-fuel technologies, over scenarios with various deployment percentages and ranges. Drop-in fuel deployments are modeled as admixtures with diesel in existing locomotives, while battery-electric deployments are shown for varying technology penetration levels and locomotive ranges. A 50% admixture ratio of diesel with biodiesel is estimated to provide a 36% emissions reduction with a cost of $0.13 per kilogram of $CO_2$ reduced, while a similar mixture with e-fuels would cut emissions by 50% at a cost of $0.22 per kilogram of $CO_2$ reduced. Battery-electric technology deployments at 50% of all ton-miles highlight the need for of technological developments in battery energy densities as scenarios for 800-mile range locomotives are estimated to provide a 46% emissions reduction at a cost of $0.06 per kilogram of $CO_2$ reduced, compared to a 16% emissions reduction at a cost of $0.11 per kilogram of $CO_2$ reduced for 400-mile range locomotives.

As the deployment of alternative fuel technologies would pose considerable operational impacts to the railroads, future work should incorporate the potential delay, congestion, additional fleet (locomotives and rail cars), and track infrastructure maintenance costs due to possible requirements and re-routing options associated with a particular deployment strategy.

Furthermore, this paper is limited in its discussion of future technological parameter uncertainties and the effects on technology roll-outs. Future work will conduct sensitivity analysis on key values for capital, operations and maintenance, locomotive and energy storage, and charging/refueling costs, technological parameters for energy storage tender car configurations, and emissions intensity values.

The presented framework provides a powerful tool for simulating, evaluating, and analyzing future alternative energy technology deployments for freight rail decarbonization. The current version of this tool is a first step in bridging the gap between scientific research and implementation of efforts to reduce the environmental impact of hard-to-decarbonize transportation modes.

## AUTHOR CONTRIBUTIONS

All authors contributed to study design, analysis and interpretation of results, and manuscript preparation. All authors reviewed the results and approved the submission of the manuscript.

## ACKNOWLEDGEMENTS

This paper is based on research funded by Advanced Research Projects Agency - Energy (ARPA-E) as part of a project titled "LOwering CO2: Models to Optimize Train Infrastructure, Vehicles, and Energy Storage (LOCOMOTIVES)" conducted by the Northwestern University Transportation Center (NUTC) in collaboration with Argonne National Laboratory (ANL). We thank all ARPA-E personnel who have provided insight and expertise that greatly assisted the research, especially Robert Ledoux and Mirjana Marden. We also thank Industry Advisory Board members and Breton Johnson who were part of the overall project and provided comments on the work. The authors remain responsible for all contents of the paper. The contents do not necessarily reflect the views of the sponsoring agency nor those of the industry advisory board members

*Hernandez, Ng, Siddique, Durango-Cohen, Elgowainy, Mahmassani, Wang, Zhou*                    268. *Compilation of State, County, and Local Anti-Idling Regulations*. Publication EPA420-B-06–004. U.S. Environmental Protection Agency, 2006, p. 102.
9. Saul, J. Amazon and Others Commit to Using Zero-Carbon Shipping Fuels by 2040. Reuters, , 2021.
10. McCall, J. B. Dieselisation of American Railroads: A Case Study. *The Journal of Transport History*, Vol. 6, No. 2, 1985, pp. 1–17. https://doi.org/10.1177/002252668500600201.
11. *Research & Investments Driving Tomorrow's Sustainable Locomotive Fleet*. American Association of Railroads, 2022.
12. *Oppose Rail Electrification and Support Sensible Climate Policy*. American Association of Railroads, 2021.
13. Duffner, F., M. Wentker, M. Greenwood, and J. Leker. Battery Cost Modeling: A Review and Directions for Future Research. *Renewable and Sustainable Energy Reviews*, Vol. 127, 2020, p. 109872. https://doi.org/10.1016/j.rser.2020.109872.
14. Popovich, N. D., D. Rajagopal, E. Tasar, and A. Phadke. Economic, Environmental and Grid-Resilience Benefits of Converting Diesel Trains to Battery-Electric. *Nature Energy*, Vol. 6, No. 11, 2021, pp. 1017–1025. https://doi.org/10.1038/s41560-021-00915-5.
15. Wang, M., A. Elgowainy, U. Lee, A. Bafana, S. Banerjee, P. T. Benavides, P. Bobba, A. Burnham, H. Cai, and U. R. Gracida-Alvarez. *Summary of Expansions and Updates in GREET® 2021*. Argonne National Lab.(ANL), Argonne, IL (United States), 2021.
16. Stephens, B. Wabtec's FLXdrive Battery-Electric Locomotive Begins Revenue Tests on BNSF. Trains, , 2021.
17. Jansen, B. *Progress Rail and Pacific Harbor Line Sign Agreement*. Progress Rail, 2020.
18. South, K. Union Pacific Railroad to Assemble World's Largest Carrier-Owned Battery-Electric Locomotive Fleet. Union Pacific News Releases, , 2022.
19. Luczak, M. CP's Hydrogen Locomotive Powers Up. Railway Age, , 2022.
20. Archetti, C., L. Peirano, and M. G. Speranza. Optimization in Multimodal Freight Transportation Problems: A Survey. *European Journal of Operational Research*, Vol. 299, No. 1, 2022, pp. 1–20. https://doi.org/10.1016/j.ejor.2021.07.031.
21. Mahmassani, H. S., K. Zhang, J. Dong, C.-C. Lu, V. C. Arcot, and E. Miller-Hooks. Dynamic Network Simulation–Assignment Platform for Multiproduct Intermodal Freight Transportation Analysis. *Transportation Research Record*, Vol. 2032, No. 1, 2007, pp. 9–16. https://doi.org/10.3141/2032-02.
22. Hwang, T., and Y. Ouyang. Assignment of Freight Shipment Demand in Congested Rail Networks. *Transportation Research Record*, Vol. 2448, No. 1, 2014, pp. 37–44. https://doi.org/10.3141/2448-05.
23. Uddin, M. M., and N. Huynh. Freight Traffic Assignment Methodology for Large-Scale Road–Rail Intermodal Networks. *Transportation Research Record: Journal of the Transportation Research Board*, Vol. 2477, No. 1, 2015, pp. 50–57. https://doi.org/10.3141/2477-06.
24. Bauer, J., T. Bektaş, and T. G. Crainic. Minimizing Greenhouse Gas Emissions in Intermodal Freight Transport: An Application to Rail Service Design. *Journal of the Operational Research Society*, Vol. 61, No. 3, 2010, pp. 530–542. https://doi.org/10.1057/jors.2009.102.
25. Nourbakhsh, S. M., and Y. Ouyang. A Structured Flexible Transit System for Low Demand Areas. *Transportation Research Part B: Methodological*, Vol. 46, No. 1, 2012, pp. 204–216. https://doi.org/10.1016/j.trb.2011.07.014.